\documentclass{kapproc} 

\usepackage[dvips]{graphicx}
\usepackage{isolatin1}

\upperandlowercase

\kluwerbib 

\begin{document}



\articletitle[{\small$z-ph$}-REM: A photometric redshift code for the 
REM telescope]{{\small$z-ph$}-REM: A photometric redshift\\ code for the 
REM telescope}

\chaptitlerunninghead{$z-ph$-REM}




\author{A. Fernández-Soto (on behalf of the REM/ROSS collaboration)}
\affil{Observatori Astronòmic de la Universitat de València, Spain}
\email{alberto.fernandez@uv.es}















\begin{abstract}
The REM telescope is being deployed these very days at the La Silla
Observatory in Chile, and will be fully operational by the beginning of
2004. It is a very fast-slewing robotized telescope, endowed with optical and
near-infrared capabilities, designed with the primary objective of performing
rapid follow-up of GRB events. One of the key issues will be the prompt
recognition of potentially interesting bursts (those happening at high
redshift or peculiarly reddened). Here I present a sketch of z-ph-REM, the
photometric redshift code designed for this mission.
\end{abstract}




\section{The GRB enigma} 
One of the major questions still under discussion about Gamma-Ray Bursts is
the fact that a large fraction (as high as 50\%) of them remains undetected
in the optical, even though facilities now exist to perform rapid, multi-site
follow-up of the events. Amongst the possible explanations for this
observational fact, the most promising ones assume it is the result of major
reddening of the source by large quantities of dust, or that the GRB sources
lie at such high redshift that the Lyman-$\alpha$ (plus Lyman-limit)
absorption by the IGM completely quenches all observed-frame visible
light. This would imply GRB redshifts $z\approx 8$ and above.

\section{The Swift mission}
Swift is a US/Italy/UK mission to be launched on 1/2004, devoted to the
detection of GRBs--an estimated 300 events/yr.  Swift includes an on-board
suite of three telescopes, including $\gamma$-rays, X-rays, and optical/UV
detectors. Once a GRB will be detected, a prealert signal shall be sent to
the control system, and the other telescopes will point to the observed
source in order to obtain an accurate position and photometry within
seconds. The information will be relayed to all linked stations.

\section{The REM telescope} 
As has just been explained, Swift will cover most of the high-energy spectrum
of the observed targets (from $\approx$ 6000 \AA, almost all the
way to $\gamma$-rays). Unfortunately, it is known that in many cases the
optical-UV telescope will show no detection. REM is a fully robotic, 60-cm,
fast-slewing, optical-near IR telescope designed to act in parallel with
Swift (but also with any other $\gamma$-ray observatory) as another
optical-infrared eye, able to point to the target position in seconds. Our
main objective is the detection and characterization of those GRBs which show
no optical counterpart, but could show as bright sources in the near infrared
range. A collaborative agreement has been signed with ESO in order to use the
REM data as trigger for the observation of such very high-redshift candidates
with larger telescopes (VLT).

\section{Photometric redshifts} 
The use of photometric redshift techniques has imposed over the last years as
a valid means for measuring redshifts in the case of: (i) many objects, (ii)
faint objects, and/or (iii) need for rapid, reasonably approximate
results. In the case of the REM targets we will use photometric redshifts in
order to estimate the redshift of the observed counterpart. Our code must
autonomously and iteratively decide whether more data are necessary, interact
with the REM control software, and send out a high-redshift candidate trigger
signal as soon as the data quality reaches the pre-defined values for a
potentially high-redshift counterpart.

The basis of the photometric redshift code is described in Lanzetta, Yahil \&
Fernández-Soto (1996, Nature, 386, 751); Fernández-Soto, Lan\-ze\-tta \&
Yahil (1999, ApJ, 513, 34); and Fernández-Soto et al. (2002, MNRAS,330,
889). It is based on a likelihood analysis that compares the observed fluxes
to the fluxes that would be measured from a series of fiducial spectra. Our
technique outputs a redshift probability function using which the quality of
the best-fit redshift value can be assessed.

One key difference between $z-ph$-REM and the codes described in the
references above is the choice of the comparison templates: $z-ph$-REM will
use a series of pure power-law spectra. The power-law
colours are preserved when the spectrum is redshifted, thus the only
detectable feature will be the one imposed by the IGM absorption. The
available selection of filters covers the whole range from 0.4 to 2.5 microns
almost continuously, making it possible to characterise GRB counterparts over
the redshift range $z \approx 2$ to $z \approx 15$.




%





\end{document}